\documentclass[12pt]{article}
\usepackage{amsfonts}
\usepackage{bbm}
\usepackage{graphicx}
\usepackage{mathrsfs}
\usepackage{color}
\usepackage{verbatim}
\usepackage{cancel}
\usepackage{ulem}
\usepackage{geometry}             
\usepackage{amssymb}
\usepackage{amsmath}
\usepackage{epstopdf}
\usepackage{cases}
\usepackage{cite}

\usepackage[hyperindex=true,
          pdfstartview=FitH,
          bookmarksnumbered=true,
          bookmarksopen=true,
          citecolor=blue,
          linkcolor=blue,
          colorlinks=true,
          urlcolor=rossoCP3,
          unicode]{hyperref}

\newcommand{\p}{\partial}

\definecolor{rossoCP3}{cmyk}{0,.88,.77,.40}

\begin{document}

\title{\bf Time evolving fluid from Vaidya spacetime}
\author{ Bin Wu${}^{1,3,4}$\thanks{{\em email}: \href{mailto:binwu@nwu.edu.cn}
{binwu@nwu.edu.cn}},
Xin Hao${}^{2}$\thanks{{\em email}: \href{mailto:shanehowe@mail.nankai.edu.cn}
{shanehowe@mail.nankai.edu.cn}}
and Liu Zhao${}^{2}$\thanks{{\em email}: \href{mailto:lzhao@nankai.edu.cn}
{lzhao@nankai.edu.cn}}\\
\vspace{-5pt}\\
\small ${}^1$Institute of Modern Physics, Northwest University, Xi’an 710069, China\\
\small ${}^2$School of Physics, Nankai University, Tianjin 300071, China\\
\small ${}^3$School of Physics, Northwest University, Xian 710069, China\\
\small ${}^4$Shaanxi Key Laboratory for Theoretical Physics Frontiers, Xi’an 710069, China\\
}

\date{}
\maketitle

\begin{abstract}
A time evolving fluid system is constructed on a timelike boundary hypersurface at finite cutoff 
in Vaidya spacetime. The approach used to construct the fluid equations is a direct extension 
of the ordinary Gravity/Fluid correspondence under the constrained fluctuation 
obeying Petrov type I conditions. The explicit relationships between the time dependent 
fluctuation modes and the fluid quantities such as density, velocity field and kinematic 
viscosity parameters are established, and the resulting fluid system is governed by a
system of a sourced continuity equation and a compressible Navier-Stokes equation with
non-trivial time evolution.
\end{abstract}

\newpage

\section{Introduction}

The AdS/CFT correspondence \cite{Maldacena:1997re,Gubser:1998bc,Witten:1998qj} 
has proven to be a powerful tool for the study of strongly coupled 
quantum field theories, especially for understanding
their transport properties \cite{Policastro:2001yc}. 
Hydrodynamics is the low-energy effective theory for slowly varying
fluctuations around thermal equilibrium of any
quantum field theory, it has been successfully
applied to describe the quark-gluon plasma created in heavy-ion collisions \cite{Kolb:2003dz}. 
Holography provides a simple prescription in which one can extract the
properties from the classical perturbation equations 
of black holes\cite{Policastro:2002se,Kovtun:2003wp}. 
It has been shown that holographic plasmas near equilibrium are close 
to perfect fluids\cite{Janik:2005zt}.

One of the important results from holographic hydrodynamics says that, 
for standard Einstein gravity with isotropic black hole horizons, the ratio of shear 
viscosity to entropy density $\eta/s$ of the holographic quark-gluon plasmas tends to take 
a universal value $\frac{1}{4\pi}$. However this value receives corrections due to 
higher curvature terms \cite{Brigante:2007nu,Brigante:2008gz,Cai:2009zv}
or in the presence of translation symmetry breaking \cite{Rebhan:2011vd,
Hartnoll:2016tri,Critelli:2014kra}. The value $\frac{1}{4\pi}$ is known as the  
Kovtun-Son-Starinets (KSS) bound \cite{Buchel:2003tz,Kovtun:2004de}. 
One utilizes the violation of KSS bound to study the relationship between
shear viscosity of boundary theories and thermodynamical phase structure
of the bulk black holes \cite{Cadoni:2017ktd,Cadoni:2017fnd}, and this promotes
our understanding to the behaviors of bulk gravity solutions.

With the success of the holographic techniques, the hydrodynamic limits of 
boundary quantum field theories have been widely investigated, and were
extend in \cite{Baier:2007ix,Bhattacharyya:2008jc} for studying
non-linear fluid dynamics and the higher order transport coefficients.
Universality of hydrodynamics also appear in the second-order transport
coefficients whose linear combination was proven to vanish both in conformal
theory duality \cite{Erdmenger:2008rm,Haack:2008xx} and non-conformal 
cases \cite{Kleinert:2016nav}.  Again this analysis is
corrected by higher curvature terms \cite{Grozdanov:2016fkt}. 

Actually, one can realize a horizon fluid which is governed 
by Damour-Navier-Stokes equation, and the seminal work could be dated back to the 
1970's \cite{Damour:1978cg}. In spite of the differences in approaches, the shear viscosity
to entropy density ratio $\eta/s$ can be evaluated in both cases
\cite{Iqbal:2008by,Eling:2009sj}. Efforts have been made 
to clarify that the Gravity/Fluid correspondence can be expressed
in membrane paradigm language by considering a fictitious membrane
at finite cutoff \cite{Son:2007vk}. In analogy to the AdS/CFT duality, 
the radial position $r_c$ of the holographic screen is related to the energy scale of 
the boundary quantum field theory, so the dependence of quantities on $r_c$ is viewed 
as the renormalization group flow in the resulting fluid
\cite{Heemskerk:2010hk,Faulkner:2010jy,Bredberg:2010ky}.

Instead of requiring an asymptotically AdS region,
the author of \cite{Bredberg:2011jq} introduced an arbitrary finite cutoff $r_c$ outside
the horizon and construct the non-relativistic dual fluid on the hypersurface.
With the Dirichlet boundary condition and the requirement of regularity at the future horizon,
it was shown that any solution of the incompressible Navier-Stokes equation could be mapped 
to a unique solution of the vacuum Einstein equation \cite{Compere:2011dx}. More importantly 
the derivative expansions and the regularity conditions
are shown to be mathematically equivalent to the near horizon expansions and
Petrov type boundary condition respectively in Gravity/Fluid correspondence. 
In this construction, it is unnecessary to solve perturbed Einstein equation, making it
mathematically much simpler and elegant \cite{Lysov:2011xx}. Numerous generalizations and 
discussions about Gravity/Fluid correspondence have been carried out along this line of 
researches \cite{cai1,wxn,HWZ,xiaoning,Huang:2011kj,Ying2,Wu:2013kqa,Ling:2013kua,WB1,WB2,
Cai:2014ywa,Pan:2016ztm}. 

Most of the existing literature concentrates on the cases where 
the bulk gravity is consisted of a stationary spacetime, and consequently the resulting 
holographic fluid possesses a stationary, non-evolving density distribution.  
It is certainly of great interests to consider fluids with non-trivial time evolution in
the framework of Gravity/Fluid correspondence. This will deepen our understanding on the 
non-equilibrium processes in the boundary field theory which, presumably may describe a
system of holographic plasma. In \cite{Bjorken:1982qr}, it is shown that 
non-equilibrium plasma behaves approximately like a perfect fluid and its stress tensor is 
determined by a single time-dependent function. The success of AdS/CFT on near-equilibrium 
cases motivates researchers to study thermalization of strongly coupled plasma. 
Ref. \cite{Janik:2006ft} investigated the dynamical time-dependent process 
in the boost invariant approximation in the context of AdS/CFT in hydrodynamics limit.
It is known that the thermalization process in the dual gauge theory
is related to the process of black hole formation in the bulk 
\cite{Chesler:2008hg,Chesler:2010bi}. The holographic entanglement entropy which was 
initially proposed in static background is also generalized to study their evolution 
during thermalize process \cite{Balasubramanian:2011ur, AbajoArrastia:2010yt}.
In the present work, we will adopt the Petrov type approach to construct a time-evolving
fluid on a finite cutoff surface in a dynamical spacetime. The simplest dynamical 
spacetime is Vaidya spacetime, which describes the formation of a Schwarzschild-AdS like 
black hole out of a collapsing shell of null dust.
According to \cite{Chesler:2008hg}, Vaidya spacetime can be used as a model 
dual to the equilibration process of the boundary theory. We find that the time-evolving
fluid living on a finite cutoff surface in Vaidya spacetime  
obeys a sourced continuity equation and a forced Navier-Stokes equation,
whose energy density and viscosity are both time-dependent. 

\section{General setup}

Consider a  $(d+2)$-dimensional Einstein gravity with a negative cosmology constant 
and a null dust as matter source. The field equation reads
\begin{align}
	R_{\mu\nu}-\frac{1}{2}R g_{\mu\nu} + \Lambda g_{\mu\nu}=\kappa T_{\mu\nu},
	\qquad \Lambda=-\frac{d(d+1)}{2\ell^2},
\end{align}
where $\ell$ represents the AdS radius and we will set $\ell$ as well as the Einstein 
constant $\kappa$ to unity in this paper.  The stress-energy tensor of null dust source reads 
\[
T_{\mu\nu}=\mu l_{\mu}l_{\nu}, \qquad l^{\mu}l_{\mu}=0.
\]
We focus ourselves on the planar Vaidya solution with the metric
\begin{align}
	\mathrm{d}s^2=-f(r,u)\mathrm{d}u^2 + 2\mathrm{d}u\mathrm{d}r 
	+ r^2\sum\limits_{I = 1}^d
	\mathrm{d}x_I^2, \qquad f(r,u)=r^2-\frac{m(u)}{r^{d-1}}, 
	 \label{metric}
\end{align}
where 
\begin{align}
	T_{uu}=\frac{d \p _u m(u)}{2r^d}
\end{align}
is the only non-vanishing component of the stress-energy tensor. 

In the study of gravitational collapse, in order to characterize
the formation of a black hole at the level of local time
evolution, there is a generalized notion
called apparent horizon which is distinct from the
event horizon. The apparent horizon is a closed spacelike hypersurface of
codimension two and is the boundary of the trapped surfaces for which the expansions 
along the two future directed normal null directions are negative. For Vaidya spacetime, the 
apparent horizon is located at
\begin{align*}
	r_h=m^{1/(d-1)}.
\end{align*}
With the null-energy condition which implies $\dot{m}(u)>0$, the radial position $r_h$ of the 
apparent horizon always grows. In contrast, the event horizon is given by
\begin{align*}
	\frac{\mathrm{d} r_e}{\mathrm{d} u}=\frac{1}{2}
	\Big(r^2-\frac{m(u)}{r^{(d-1)}}\Big),
	\qquad \lim\limits_{u\rightarrow\infty}r_e
	=m^{1/(d-1)}.
\end{align*}
We will assume that the mass
function $m(u)$ tends to a finite value $M$ 
as $u\rightarrow\infty$, so that the final state of the spacetime (\ref{metric}) 
corresponds to the geometry of Schwarzschild black hole with a planar 
topology. In the timelike asymptotic region, it is known that the apparent horizon and event
horizon coincide in static spherical spacetime.

The physically meaningful profile for the mass function $m(u)$ should be increasing from $0$ 
to a finite value $M$, which may be chosen to be
\begin{align}
	m(u)=\frac{M}{2}(\mathrm{tanh}(u/v_0)+1). \label{mass}
\end{align}
This profile describes a geometry transition from a pure planar AdS spacetime to a 
Schwarzschild-AdS black brane. The parameter $v_0$ characterizes the thickness of the null 
dust shell which falls along $u=0$. When $v_0$ tend to zero, the mass
function becomes to a step function $M\theta(u)$. 

To construct holographic fluid living on the codimension-1 
timelike hypersurface, we will always keep the intrinsic metric of the hypersurface fixed 
since we are mainly interested in the evolution of dual fluid
caused by the process of the formation of bulk black hole. As a consequence, the near horizon 
limit which is usually taken in Petrov type I approach of Gravity/Fluid correspondence is 
not valid any more, because of the time dependences of both the event and apparent horizons. 
For this reason we take the holographic screen $\Sigma_c$ at some finite cutoff $r=r_c$ 
outside the apparent horizon. In view of the renormalization group interpretation,
the radial direction corresponds to energy scale of the boundary theory, so the finite cutoff 
implies a finite energy scale, which is physically more realistic and may be reached by
experiment. 

\section{Constraints on the hypersurface}

The geometry of the hypersurface $\Sigma_c$ is best characterized by its first and second 
fundamental forms. The first fundamental form is the projection tensor
\begin{align}
 	h_{\mu\nu}=g_{\mu\nu}-n_{\mu}n_\nu,
 	\label{ndlelm}
\end{align}
where 
\begin{align}
	n_\mu=\Big(0,\frac{1}{\sqrt{f}},0,0\Big), \,\,\,
	n^\mu=\Big(\frac{1}{\sqrt{f}},\sqrt{f},0,0\Big), 
\end{align}
is a unit spacelike normal vector with $n_{\mu}n^{\mu}=1$. Evidently
one can identify $h_{\mu\nu}$ with the induced metric on $\Sigma_c$. More precisely, 
the induced metric $\gamma_{ab}$ on $\Sigma_c$ can be written as
\[
\gamma_{ab} = h_{\mu\nu} e^\mu{}_a e^\nu{}_b,
\]
where $e^\mu{}_a = \frac{\partial x^\mu}{\partial y^a}$ with $x^\mu = (u, r, x^I)$ and 
$y^a = (u, x^I)$. The second fundamental form is the extrinsic curvature
\begin{align}
	K_{\mu\nu}=\frac{1}{2}\mathscr{L}_n h_{\mu\nu},
\end{align} 
whose perturbations will be considered as the fundamental
variables in the dual theory.

Projecting the field equations in the transverse and longitudinal directions, 
we get so-called a Hamiltonian and momentum constraints
\begin{align*}
&\hat{R} + K^{ab} K_{ab} - K^2 = 2\Lambda-2T_{\mu\nu}n^{\mu}n^{\nu}, \\
&D_a (K^a{}_b - \gamma^a{}_b K) = T_{\mu\nu}n^{\mu}h^{\nu}{}_b, 
\end{align*}
where $\hat{R}$ is the Ricci scalar on the hypersurface $\Sigma_c$, $D_a$ is the
covariant derivative compatible with induced metric. These 
constraints play essential roles in the construction of the fluid equations.

The definition of Brown-York stress tensor is 
\begin{align}
	t_{ab}=h_{ab}K-K_{ab}-Ch_{ab},   \label{B-Y}
\end{align}
where the last term is a counter term to remove the divergence if the cutoff
is taken at infinity. Here we take $C=0$ since we consider only a finite cutoff.
The Brown-York stress tensor is regarded as the stress-energy tensor of the dual fluid 
and mapped into polynomials in the extrinsic curvature through (\ref{B-Y}) 
on the boundary hypersurface. Therefore, we can take $t_{ab}$, instead of $K_{ab}$, 
as the fundamental dynamical quantity in the boundary theory.
Then constraint equations are reformulated as
\begin{align}
& \mathscr{H} = \hat{R} + t^a{}_b t^b{}_a
- \frac{t^2}{2} = 2\Lambda-2T_{\mu\nu}n^{\mu}n^{\nu},   \label{hamc}     \\
& \mathscr{P}_b = D_a t^a{}_b = T_{\mu\nu}n^{\mu}h^{\nu}{}_b,      \label{momc}
\end{align}
where $t=\gamma^{ab}t_{ab}$ is the trace of the Brown-York tensor.

Petrov type is a seminal proposal for classifying spacetime with certain symmetries.
The Vaidya spacetime (\ref{metric}) we are working in is of Petrov type D, which also belongs 
to Petrov type I. Taking the Vaidya spacetime (\ref{metric}) as the initial data and demanding 
that the spacetime evolves no singularity after perturbation, the geometry of the
perturbed spacetime in the vicinity of the hypersurface $\Sigma_c$ 
should at least be of Petrov type I \cite{Lysov:2011xx}. 
Thus we get the following the additional constraints
\begin{align}
P_{IJ} =
l^\mu(m_I )^\nu l^\sigma(m_J)^\rho
C_{\mu\nu\rho\sigma} \big|_{\Sigma_c} = 0,
\label{PTI}
\end{align}
where $C_{\mu\nu\rho\sigma}$ is the Weyl tensor of the bulk spacetime and
the Newman-Penrose-like basis vector fields $l^\mu, k^\mu, (m_I)^\mu$ are introduced, 
which obey
\begin{align}
	&  l^2=k^2=0,\,(k,l)=1,\,(l,m_{i})=(k,m_{I})=0,\,(m_{I},m_{J})=\delta^{I}{ }_{J}.
\end{align}
In coordinates system $x^\mu = (u,r, x_{I})$, the basis vector fields restricted on the hypersurface $\Sigma_c$ read
\begin{align*}
	& m_{I}=\frac{1}{r_c} \partial_{J},\\
	& l=\frac{1}{\sqrt{2}}\left(\frac{1}{\sqrt{f_c}}\partial_{u}-n\right),\\
	& k=-\frac{1}{\sqrt{2}}\left(\frac{1}{\sqrt{f_c}}\partial_{u}+n\right),
\end{align*}
where we have used the notation $f_c =f_c(u)= f(r_c,u)$ for short. 
Inserting N-P basises into Petrov type condition (\ref{PTI}),
we get
\begin{align*}
 	P_{IJ}= \frac{1}{f_c} C_{uIuJ}+ \frac{1}{\sqrt{f_c}} C_{uIJ(n)}+ 
 	\frac{1}{\sqrt{f_c}} C_{uJI(n)}+C_{I(n)J(n)}=0. 
\end{align*}
The symmetric Petrov type I condition provides $\frac{d(d+1)}{2}-1$ constraints
over Brown-York stress tensor, where the extra $-1$ is provided by the 
traceless condition for the Weyl tensor. Therefore the number of independent components for the 
Brown-York stress tensor reduces from $\frac{(d+1)(d+2)}{2}$ to $d+2$, 
which is exactly the number of variables required to describe the energy density, pressure 
and velocity fields of a holographic fluid system. 
In the presence of matter source in the bulk, there are some extra degrees of freedom
for the matter field, which will play the role of
fluid source or external forces \cite{Bhattacharyya:2008ji,Cai:2012vr}.

\section{Non-relativistic hydrodynamic expansion}

In order to construct the explicit fluid equations on the finite cutoff surface $\Sigma_c$, 
there remains one more step which is the hydrodynamic expansion of the perturbed 
constraint equations. In static background spacetimes, this last step is often performed
in the near horizon limit. In view of the RG group flow approach, near horizon limit
implies only the low energy modes of the perturbations contributes to the 
hydrodynamics description of the boundary system. However, in our case, we cannot do this 
because of the evolution of the horizons. Instead, what we need is to take the long wavelength
limit on the finite cutoff surface, which involves contributions from degrees of freedom 
at relatively bigger energy scale whilst still maintaining the validity of the hydrodynamic
explanation.

The line element corresponding to the induced metric on $\Sigma_c$ reads
\begin{align}
 	\mathrm{d} s_{ d +1}^2 &= -f_c(u) \mathrm{d} u^2++ r_c^2\sum\limits_{I = 1}^d
	\mathrm{d}x_I^2,  \nonumber \\
 	&=-\frac{1}{\lambda^2}f_c(u(\tau))\mathrm{d}\tau^2 +\frac{1}{\lambda}
 	\sum\limits_{i = 1}^d
	\mathrm{d}x_i^2,
 	\label{induced}
\end{align}
where in the second line we have introduced a non-relativistic rescaling $\tau=\lambda u$, 
$x_i=\sqrt{\lambda}r_c x_I$, where $\lambda$ is taken to be a small dimensionless parameter. 
The effect of the rescaling $\tau=\lambda u$ sharpens the mass function \eqref{mass}, 
or effectively speeds up the formation of the black hole, but this will not affect the 
asymptotic behavior of the function $f(r,u)$. The different scalings in $u$ and $x_I$ allows us
to interpret the limit $\lambda\to 0$ as both the non-relativistic and the long wavelength 
limit.

Now in coordinates $(\tau,x_i)$, we get the non-relativistic hydrodynamic expansion of 
the Hamiltonian constraint
\begin{align}
	\mathcal{H}=(t^\tau{}_\tau)^2- 2t^\tau{}_i\frac{f_c}{\lambda^2} 
	t^\tau{}_j h^i{}_j +t^i{}_jt^j{}_i
	-\frac{t^2}{d}-2\Lambda-2T_{\mu\nu}n^{\mu}n^\nu,   \label{Hac}
\end{align}
as well as Petrov type I condition in terms of the Brown-York stress tensor,
\begin{align}
	P_{ij}&= \frac{2f_c}{\lambda^2}t^{\tau}{ }_{i}t^{\tau}{ }_{j} + \frac{t^2}{d^2}h_{ij}
	-\frac{t}{d}t^{\tau}{ }_{\tau}h_{ij} + t^{\tau}{ }_{\tau}t_{ij} 
	+ \frac{2\lambda}{\sqrt{f_c}}\partial_{\tau}\left(\frac{t}{d}h_{ij}-t_{ij}\right) 
	\nonumber\\
	&\quad - \frac{2\sqrt{f_c}}{\lambda}\p_{{(}i}t^{\tau}{ }_{j{)}}- t_{ik}t^{k}{ }_{j}     
 	+A_{ij}=0,     \label{PTc}
\end{align}
where $A_{ij}$ represents the following contribution from the bulk matter and the cosmological 
constant,
\begin{align}
   A_{ij}=-\frac{1}{d}(T_{\nu\rho}n^{\nu}n^{\rho}-2\Lambda+T+\frac{1}{f}T_{uu}
   -\frac{2}{\sqrt{f}}T_{u\rho}n^{\rho})h_{ij}+T_{ij}.  \label{matter}
\end{align}

On the background level, the Brown-York tensor on the finite cutoff surface
can be evaluated explicitly, yielding
\begin{align*}
	&t^{\tau}{}_{\tau}=\frac{d\sqrt{f_c}}{r},  \qquad \qquad
	t^{\tau}{ }_{i}=0, \\
	&t^{i}{}_{j}=\left(\frac{1}{2\sqrt{f_c}}\partial_{r_c}f
	+\frac{(d-1)\sqrt{f_c}}{r_c}- \frac{\lambda\partial_\tau f_c}{2f_c^{3/2}}\right)h^{i}{ }_{j},   \\
	&t=d\bigg(\frac{1}{2\sqrt{f_c}}\partial_{r_c}{f}+ \frac{d \sqrt{f_c}}{r}
	-\frac{\lambda\partial_\tau f_c}{2f_c^{3/2}}\bigg).
\end{align*}
All the constraint equations are automatically satisfied on the background level, 
since these equations are actually the field equations based on which we deduce the 
background solutions. As mentioned earlier, what matters in the hydrodynamic description is 
the long wavelength perturbation modes of the Brown-York tensor. So it necessary to 
consider expansions around the background values. The most general fluctuations can be very 
complicated, for simplicity, we restrict ourselves to the following 
polynomial expansion, 
\begin{align}
	t^{a}{ }_{b}=\sum^{\infty}_{n=1} \lambda^{n}t^{a}{ }_{b}^{(n)},
\end{align}
where we have intentionally taken the expansion parameter to be 
identical to the scaling parameter $\lambda$, so that the perturbative limit 
$\lambda \rightarrow 0$ is simultaneously the non-relativistic
and long wavelength limit.

After some tedious calculations we find that the perturbed Hamiltonian constraint yields, at 
the first nontrivial order $O(\lambda)$, the following equation,
\begin{align}
    {t^{\tau}}_{\tau}^{(1)}
    &=\frac{2r_c f_c^{2/3}}{-r_c\partial_{r_c}f+2 f_c}
     \delta^{ij}t^{\tau}{}_i^{(1)}t^{\tau}{}_j^{(1)}
     +\frac{2f_c}{-r_c\partial_{r_c} f+2 f_c}t^{(1)}  \nonumber \\
    &\quad +\frac{d(r\partial_{r_c} f +2(d-2) f_c)\partial_\tau f_c}
    {2(-r\partial_{r_c} f+2 f_c)f_c^{3/2}}
     +\frac{2T_{\mu\nu}n^{\mu}n^{\nu}
     r\sqrt{f_c}}{-r\partial_{r_c} f+2 f_c}.  \label{hacc}
\end{align}
Meanwhile, the perturbed Petrov type I condition at order $\mathcal{O}(\lambda)$ yields
\begin{eqnarray}
    t^{i}{}_j^{(1)} &=& \frac{2r_cf_c^{3/2}}{r_c\partial_{r_c}f
    +(d-2)f_c}{{t^{\tau}}_k}^{(1)}{{t^{\tau}}_j}^{(1)}\delta^{ik}
    -\frac{2r_cf}{r_c\partial_{r_c}f+(d-2)f_c}
    \p_{(k}{t^{\tau}{}_{j)}}^{(1)}\delta^{ik} \nonumber\\
     &&-\frac{f_c}{r_c\partial_{r_c}f+(d-2)f_c}
    {\delta^i}_j{{t^{\tau}}_{\tau}}^{(1)}+\frac{r_c\partial_{r_c}f+df_c}
    {r_c\partial_{r_c}f+(d-2)f_c}\frac{t^{(1)}}{d}{\delta^i}_j. \label{PTcc}
\end{eqnarray}
Let us remind that apart from the contributions from the matter source and the background level
metric functions, the last two equations are consisted purely of the first order perturbations
of the Brown-York tensor. We will see in the next section that, by introducing an
appropriate holographic dictionary, the momentum constraint imposed on $t^{a}{}_b^{(1)}$ 
can be interpreted as the sourced continuity equation and the Navier-Stokes equation
for a compressible fluid living on the finite cutoff surface $\Sigma_c$, which subjects to a 
deviatoric stress and an external force term.

\section{Holographic fluid in Vaidya spacetime} 
 
Now let us consider the momentum constraint imposed on the cutoff hypersurface $\Sigma_c$. 
Since the hypersurface is flat, the covariant divergence appearing in \eqref{momc} can 
actually be rewritten as ordinary divergence, thus we have
\begin{align}
\p_{a}t^{a}{ }_{b}=-T_{\mu b} n^{\mu}. \label{divnf}
\end{align}

Inserting the relations (\ref{hacc}) and (\ref{PTcc})
from the Hamiltonian constraint and Petrov type I condition into (\ref{divnf}) 
and introducing the following holographic dictionary,
\begin{align*}
	&t^\tau{}_i^{(1)}= \dfrac{r_c\partial_{r_c} f +(d-2)f_c}{2r_cf_c^{(2/3)}}  v_i,   
	\\
	&P=-\frac{f_c}{-r_c\partial_{r_c} f+2f_c}v_kv_l\delta^{kl} 
	+\frac{-2r_c^2f_c^{3/2}\partial_{r_c} f}{d(r_c\partial_{r_c} f +(d-2)f_c)
	(-r_c\partial_{r_c} f +2 f_c)}t^{(1)},  \\
\end{align*}
where $v_i$ and $P$ are to be understood as the velocity field and the pressure of the 
dual fluid respectively, the temporal component of momentum constraints 
at order $\mathcal{O}(\lambda^0)$ becomes a sourced continuity equation
\begin{align*}
\p_\tau \rho + \rho \delta^{ij}  \p_i v_j = Q,
\end{align*}
while the spatial components turn out to be
the Navier-Stokes equation at order $\mathcal{O}(\lambda^1)$:
\begin{align*}
	\p_\tau v_i + v^j \p_j v_i= -\frac{1}{\rho}\p_i P
	+ \nu \p^j d_{ij} + f_i,
\end{align*}
where the symmetric traceless tensor
\begin{align}
	d_{ij}= \p_j v_i + \p_i v_j -\frac{2}{d}\delta_{ij}\p^k v_k,
\end{align}
represents the deviatoric stress, which depends only on the 
first derivatives of velocity field, and 
\begin{align*}
	&\rho=\dfrac{r_c\partial_{r_c} f +(d-2)f_c}{2r_c f_c^{3/2}},  \\
	&Q=\frac{1}{f_c^{3/2}}\big(r_c\p_{r_c} - \frac{3\p_{r_c} f}{2f_c}
	+\frac{d+2}{4r_c}\big)\p_\tau f_c,  \\
	&f_i=-\frac{Q}{\rho}v_i,
\end{align*}
represent energy density, source in continuity equation and a body force acting on the fluid 
system. It is clear that the body force could 
be recognized as a linear resistance force since it linearly 
proportional to the velocity field $v_i$.
The kinematic viscosity $\nu$ of the fluid is determined solely by the background level
metric function
\begin{align}
	&\nu=\frac{r_cf_c}{r_c\partial_{r_c} f +(d-2)f_c}.  \label{visc}
\end{align}
Since the metric function $f_c(\tau)$ is time-dependent (no dependence
on boundary spatial coordinates), the energy density as well as the kinematic viscosity 
in this dual fluid system are both time dependent, and hence the continuity equation implies 
compressibility rather than incompressibility as one often finds in static background 
spacetimes. 

In the late time limit $\tau\to \infty$, the mass function $m(u(\tau))$ becomes a constant
$M$, the corresponding background spacetime becomes static, which is the final state of the 
collapsing matter. In this limit, the density of the holographic fluid becomes constant and the 
continuity equation becomes that of an incompressible unsourced fluid.
Meanwhile, one can see that the external force term $f_i$ vanishes in this limit. 
However, comparing the kinematic viscosity \eqref{visc} in the late time limit with the result 
obtained in \cite{Ling:2013kua} for the static case, which reads
\begin{align}
\nu=\frac{r_c\sqrt{f_c}}{r_c\partial_{r_c} f +(d-2)f_c}, \label{visc2}
\end{align}
one can find that there is a slight difference between the two results. 
The reason behind this difference lies in that the induced metric (\ref{induced})
we choose contains an extra time-scaling factor $f_c(\tau)$. In principle,
this factor can be absorbed by a second rescaling of the timelike coordinate via
$\sqrt{f_c(\tau)} \mathrm{d}\tau= \mathrm{d}t$. Performing this second rescaling we get
\begin{align*}
	\p_\tau \rightarrow \sqrt{f_c}\p_t, \qquad    v_i\rightarrow \sqrt{f_c}v_i.  
\end{align*}
Therefore, reformulating the resulting Navier-Stokes equation in terms of the new time 
coordinate $t$ we will get the new kinematic viscosity as given in \eqref{visc2}.
It should be emphasized that the second time rescaling can only be done in the temporal 
asymptotics $u\rightarrow  \pm\infty$ without introducing extra complexities.   
In the process of the collapse, there is no obvious reason of doing the second time rescaling  
and so we should stick to \eqref{visc} as the expression for kinematic viscosity.

The thermodynamics of dynamical spacetimes is still an open question,
since there are some difficulties in defining thermodynamic quantities like temperature, 
entropy etc. which are based on equilibrium state of system. Therefore, 
kinematic viscosity to entropy density ratio in our case remains unclear. 
A possible way to solve this problem is to consider the proposal made in \cite{Hayward:1993ph,
Hu:2015xva}, i.e. making use of the unified first law derived from the field equation 
which is also valid on apparent horizons. 

\section{Conclusion}

Motivated by studies on non-equilibrium problems
of dual field theory within the context of AdS/CFT, 
we constructed the Gravity/Fluid correspondence in time-dependent
dynamical bulk spacetime. The Vaidya geometry as the simplest lab
has been adopted, which describes the process of 
formation of a Schwarzschild-AdS black hole in AdS spacetime.

On a fixed, timelike hypersurface at finite cutoff, 
the second fundamental form, i.e. extrinsic curvature is regarded as
the fundamental quantity of the dual theory. An equivalent set of variables, i.e. the 
Brown-York stress-energy tensor, may be chosen to describe the dual fluid. Under the 
Petrov type I conditions, it is shown that the remaining unconstrained independent
components of the Brown-York stress-energy tensor obey a system of equations which can be 
identified to the continuity equation and the compressible Navier-Stokes equation.

According to the AdS/CFT, the non-equilibrium phenomenon
of quark-gluon plasma can be described by dual gravity geometry. 
In this paper, we also obtained the time-dependent 
energy density and kinematic viscosity of the non-relativistic dual fluid
by using Petrov type approach in the context of Gravity/Fluid
correspondence. Unlike usual works with static backgrounds, where
the energy densities are either constants which naturally lead to 
incompressibility conditions or hypersurface spatial coordinate dependent 
due to anisotropy \cite{Moskalets:2014hoa,Hao:2015pal}, 
which correspond to compressible but stationary fluids,
in the present case, the density of the fluid evolves with time 
and the continuity equation indicates that the fluid is both compressible and sourced. 
Meanwhile, the kinematic viscosity is also dependent on both time coordinate and the 
radial cutoff $r_c$. The dependence on time is inherited 
from the dynamical time-dependent bulk spacetime,
and the $r_c$-dependence should be considered as an effective energy scale. 
The Navier-Stokes equation also receives a time-dependent external force density which
can be regarded as linear resistance force.

\section*{Acknowledgment}

This work is supported by the National Natural Science Foundation of China under Grant
No. 11575088 and No. 11605137.

\providecommand{\href}[2]{#2}\begingroup
\footnotesize\itemsep=0pt
\providecommand{\eprint}[2][]{\href{http://arxiv.org/abs/#2}{arXiv:#2}}
\providecommand{\doi}[2]{\href{http://dx.doi.org/#2}{#1}}

\endgroup


\begin{thebibliography}{99}
\bibitem{Maldacena:1997re} 
  J.~M.~Maldacena,
  ``The Large N limit of superconformal field theories and supergravity,''
  \doi{Adv.\ Theor.\ Math.\ Phys.\  {\bf 2}, 231 (1998)}
  {doi:10.1023/A:1026654312961}
  [\eprint{hep-th/9711200}].

\bibitem{Gubser:1998bc} 
  S.~S.~Gubser, I.~R.~Klebanov and A.~M.~Polyakov,
  ``Gauge theory correlators from noncritical string theory,''
  \doi{Phys.\ Lett.\ B {\bf 428}, 105 (1998)}
  {doi:10.1016/S0370-2693(98)00377-3}
  [\eprint{hep-th/9802109}].

\bibitem{Witten:1998qj} 
  E.~Witten,
  ``Anti-de Sitter space and holography,''
  Adv.\ Theor.\ Math.\ Phys.\  {\bf 2}, 253 (1998)
  [\eprint{hep-th/9802150}].

\bibitem{Policastro:2001yc} 
  G.~Policastro, D.~T.~Son and A.~O.~Starinets,
  ``The Shear viscosity of strongly coupled N=4 supersymmetric Yang-Mills plasma,''
  \doi{Phys.\ Rev.\ Lett.\  {\bf 87}, 081601 (2001)}
  {doi:10.1103/PhysRevLett.87.081601}
  [\eprint{hep-th/0104066}].

\bibitem{Kolb:2003dz} 
  P.~F.~Kolb and U.~W.~Heinz,
  ``Hydrodynamic description of ultrarelativistic heavy ion collisions,''
  Quark gluon plasma 634-714
  [\eprint{nucl-th/0305084}].

\bibitem{Policastro:2002se} 
  G.~Policastro, D.~T.~Son and A.~O.~Starinets,
  ``From AdS / CFT correspondence to hydrodynamics,''
  \doi{JHEP {\bf 0209}, 043 (2002)}
  {doi:10.1088/1126-6708/2002/09/043}
  [\eprint{hep-th/0205052}].

\bibitem{Kovtun:2003wp} 
  P.~Kovtun, D.~T.~Son and A.~O.~Starinets,
  ``Holography and hydrodynamics: Diffusion on stretched horizons,''
  \doi{JHEP {\bf 0310}, 064 (2003)}
  {doi:10.1088/1126-6708/2003/10/064}
  [\eprint{hep-th/0309213}].


\bibitem{Janik:2005zt} 
  R.~A.~Janik and R.~B.~Peschanski,
  ``Asymptotic perfect fluid dynamics as a consequence of Ads/CFT,''
  \doi{Phys.\ Rev.\ D {\bf 73}, 045013 (2006)}
  {doi:10.1103/PhysRevD.73.045013}
  [\eprint{hep-th/0512162}].

\bibitem{Buchel:2003tz} 
  A.~Buchel and J.~T.~Liu,
  ``Universality of the shear viscosity in supergravity,''
  \doi{Phys.\ Rev.\ Lett.\  {\bf 93}, 090602 (2004)}
  {doi:10.1103/PhysRevLett.93.090602}
  [\eprint{hep-th/0311175}].

\bibitem{Kovtun:2004de} 
  P.~Kovtun, D.~T.~Son and A.~O.~Starinets,
  ``Viscosity in strongly interacting quantum field theories from black hole physics,''
  \doi{Phys.\ Rev.\ Lett.\  {\bf 94}, 111601 (2005)}
  {doi:10.1103/PhysRevLett.94.111601}
  [\eprint{hep-th/0405231}].

\bibitem{Brigante:2007nu} 
  M.~Brigante, H.~Liu, R.~C.~Myers, S.~Shenker and S.~Yaida,
  ``Viscosity Bound Violation in Higher Derivative Gravity,''
  \href{http://dx.doi.org/doi:10.1103/PhysRevD.77.126006}
  {Phys.\ Rev.\ D {\bf 77}, 126006 (2008)}
  [\eprint{0712.0805}].

\bibitem{Brigante:2008gz} 
  M.~Brigante, H.~Liu, R.~C.~Myers, S.~Shenker and S.~Yaida,
  ``The Viscosity Bound and Causality Violation,''
  \doi{Phys.\ Rev.\ Lett.\  {\bf 100}, 191601 (2008)}
  {doi:10.1103/PhysRevLett.100.191601}
  [\eprint{0802.3318}].

\bibitem{Cai:2009zv} 
  R.~G.~Cai, Z.~Y.~Nie, N.~Ohta and Y.~W.~Sun,
  ``Shear Viscosity from Gauss-Bonnet Gravity with a Dilaton Coupling,''
  \doi{Phys.\ Rev.\ D {\bf 79}, 066004 (2009)}
  {doi:10.1103/PhysRevD.79.066004}
  [\eprint{0901.1421}].

\bibitem{Rebhan:2011vd} 
  A.~Rebhan and D.~Steineder,
  ``Violation of the Holographic Viscosity Bound in a Strongly Coupled Anisotropic Plasma,''
  \doi{Phys.\ Rev.\ Lett.\  {\bf 108}, 021601 (2012)}
  {doi:10.1103/PhysRevLett.108.021601}
  [\eprint{1110.6825}].

\bibitem{Hartnoll:2016tri} 
  S.~A.~Hartnoll, D.~M.~Ramirez and J.~E.~Santos,
  ``Entropy production, viscosity bounds and bumpy black holes,''
  \doi{JHEP {\bf 1603}, 170 (2016)}
  {doi:10.1007/JHEP03(2016)170}
  [\eprint{1601.02757}].

\bibitem{Critelli:2014kra} 
  R.~Critelli, S.~I.~Finazzo, M.~Zaniboni and J.~Noronha,
  ``Anisotropic shear viscosity of a strongly coupled non-Abelian plasma from magnetic branes,''
  \doi{Phys.\ Rev.\ D {\bf 90}, no. 6, 066006 (2014)}
  {doi:10.1103/PhysRevD.90.066006}
  [\eprint{1406.6019}].

\bibitem{Cadoni:2017ktd} 
  M.~Cadoni, E.~Franzin and M.~Tuveri,
  ``Hysteresis in $\eta/s$ for QFTs dual to spherical black holes,''
  [\eprint{1703.05162}].

\bibitem{Cadoni:2017fnd} 
  M.~Cadoni, E.~Franzin and M.~Tuveri,
  ``Van der Waals-like Behaviour of Charged Black Holes and Hysteresis in the Dual QFTs,''
  \doi{Phys.\ Lett.\ B {\bf 768}, 393 (2017)}
  {doi:10.1016/j.physletb.2017.02.060}
  [\eprint{1702.08341}].

\bibitem{Baier:2007ix} 
  R.~Baier, P.~Romatschke, D.~T.~Son, A.~O.~Starinets and M.~A.~Stephanov,
  ``Relativistic viscous hydrodynamics, conformal invariance, and holography,''
  \doi{JHEP {\bf 0804}, 100 (2008)}
  {doi:10.1088/1126-6708/2008/04/100}
  [\eprint{0712.2451}].

\bibitem{Bhattacharyya:2008jc} 
  S.~Bhattacharyya, V.~E.~Hubeny, S.~Minwalla and M.~Rangamani,
  ``Nonlinear Fluid Dynamics from Gravity,''
  \doi{JHEP {\bf 0802}, 045 (2008)}
  {doi:10.1088/1126-6708/2008/02/045}
  [\eprint{0712.2456}].

\bibitem{Erdmenger:2008rm} 
  J.~Erdmenger, M.~Haack, M.~Kaminski and A.~Yarom,
  ``Fluid dynamics of R-charged black holes,''
  \doi{JHEP {\bf 0901}, 055 (2009)}
  {doi:10.1088/1126-6708/2009/01/055}
  [\eprint{0809.2488}].

\bibitem{Haack:2008xx} 
  M.~Haack and A.~Yarom,
  ``Universality of second order transport coefficients from the gauge-string duality,''
  \doi{Nucl.\ Phys.\ B {\bf 813}, 140 (2009)}
  {doi:10.1016/j.nuclphysb.2008.12.028}
  [\eprint{0811.1794}].

\bibitem{Kleinert:2016nav} 
  P.~Kleinert and J.~Probst,
  ``Second-Order Hydrodynamics and Universality in Non-Conformal Holographic Fluids,''
  \doi{JHEP {\bf 1612}, 091 (2016)}
  {doi:10.1007/JHEP12(2016)091}
  [\eprint{1610.01081}].

\bibitem{Grozdanov:2016fkt} 
  S.~Grozdanov and A.~O.~Starinets,
  ``Second-order transport, quasinormal modes and zero-viscosity limit in the Gauss-Bonnet holographic fluid,''
  \doi{JHEP {\bf 1703}, 166 (2017)}
  {doi:10.1007/JHEP03(2017)166}
  [\eprint{1611.07053}].

\bibitem{Damour:1978cg} 
  T.~Damour,
  ``Black Hole Eddy Currents,''
  \doi{Phys.\ Rev.\ D {\bf 18}, 3598 (1978)}
  {doi:10.1103/PhysRevD.18.3598}

\bibitem{Iqbal:2008by} 
  N.~Iqbal and H.~Liu,
  ``Universality of the hydrodynamic limit in AdS/CFT and the membrane paradigm,''
  \doi{Phys.\ Rev.\ D {\bf 79}, 025023 (2009)}
  {doi:10.1103/PhysRevD.79.025023}
  [\eprint{0809.3808}].

\bibitem{Eling:2009sj} 
  C.~Eling and Y.~Oz,
  ``Relativistic CFT Hydrodynamics from the Membrane Paradigm,''
  \doi{JHEP {\bf 1002}, 069 (2010)}
  {doi:10.1007/JHEP02(2010)069}
  [\eprint{0906.4999}].

\bibitem{Son:2007vk} 
  D.~T.~Son and A.~O.~Starinets,
  ``Viscosity, Black Holes, and Quantum Field Theory,''
  \doi{Ann.\ Rev.\ Nucl.\ Part.\ Sci.\  {\bf 57}, 95 (2007)}
  {doi:10.1146/annurev.nucl.57.090506.123120}
  [\eprint{0704.0240}].

\bibitem{Heemskerk:2010hk} 
  I.~Heemskerk and J.~Polchinski,
  ``Holographic and Wilsonian Renormalization Groups,''
  \doi{JHEP {\bf 1106}, 031 (2011)}
  {doi:10.1007/JHEP06(2011)031}
  [\eprint{1010.1264}].

\bibitem{Faulkner:2010jy} 
  T.~Faulkner, H.~Liu and M.~Rangamani,
  ``Integrating out geometry: Holographic Wilsonian RG and the membrane paradigm,''
  \doi{JHEP {\bf 1108}, 051 (2011)}
  {doi:10.1007/JHEP08(2011)051}
  [\eprint{1010.4036}].

\bibitem{Bredberg:2010ky} 
  I.~Bredberg, C.~Keeler, V.~Lysov and A.~Strominger,
  ``Wilsonian Approach to Fluid/Gravity Duality,''
  \doi{JHEP {\bf 1103}, 141 (2011)}
  {doi:10.1007/JHEP03(2011)141}
  [\eprint{1006.1902}].

\bibitem{Bredberg:2011jq} 
  I.~Bredberg, C.~Keeler, V.~Lysov and A.~Strominger,
  ``From Navier-Stokes To Einstein,''
  \doi{JHEP {\bf 1207}, 146 (2012)}
  {doi:10.1007/JHEP07(2012)146}
  [\eprint{1101.2451}].

\bibitem{Compere:2011dx} 
  G.~Compere, P.~McFadden, K.~Skenderis and M.~Taylor,
  ``The Holographic fluid dual to vacuum Einstein gravity,''
  \doi{JHEP {\bf 1107}, 050 (2011)}
  {doi:10.1007/JHEP07(2011)050}
  [\eprint{1103.3022}].


\bibitem{Lysov:2011xx} 
  V.~Lysov and A.~Strominger,
  ``From Petrov-Einstein to Navier-Stokes,''
  [\eprint{1104.5502}].

\bibitem{cai1}
	R.~G.~Cai, L.~Li and Y.~L.~Zhang, 
	``Non-Relativistic fluid dual to
	asymptotically AdS gravity at finite cutoff surface'',  
	\doi{JHEP {\bf 1107}, 027 (2011)}
  {doi:10.1007/JHEP07(2011)027}
	[\eprint{1104.3281}].

\bibitem{xiaoning}
	T.~Z.~Huang, Y.~Ling, W.~J.~Pan, Y.~Tian, and X.~N.~Wu, 
	``From Petrov-Einstein to Navier-Stokes in spatially curved spacetime'',
	\doi{JHEP {\bf 1110}, 079 (2011)}
  {doi:10.1007/JHEP10(2011)079}
	[\eprint{1107.1464}].


\bibitem{Huang:2011kj}
  T.~Z.~Huang, Y.~Ling, W.~J.~Pan, Y.~Tian and X.~N.~Wu,
  ``Fluid/gravity duality with Petrov-like boundary condition in a spacetime
  with a cosmological constant'',
  \doi{Phys.\ Rev.\ D {\bf 85}, 123531 (2012)}
  {doi:10.1103/PhysRevD.85.123531}
  [\eprint{1111.1576}].

\bibitem{Ying2}
	C.~Y.~Zhang, Y.~Ling, C.~Niu, Y.~Tian, X.~N.~Wu, 
	``Magnetohydrodynamics from gravity'', 
	\doi{Phys.\ Rev.\ D {\bf 86}, 084043 (2012)}
  {doi:10.1103/PhysRevD.86.084043 }
	[\eprint{1204.0959}].

\bibitem{Wu:2013kqa}
  X.~Wu, Y.~Ling, Y.~Tian and C.~Zhang,
  ``Fluid/Gravity Correspondence for General Non-rotating Black Holes'',
	\doi{Class.\ Quant.\ Grav.\  {\bf 30}, 145012 (2013)}
  {doi:10.1088/0264-9381/30/14/145012}
  [\eprint{1303.3736}].

\bibitem{WB1} 
	B.~Wu, L.~Zhao, 
	``Gravity mediated holography in fluid dynamics'', 
	\doi{Nucl.\ Phys.\ B {\bf 874}, 177 (2013)}
  {doi:10.1016/j.nuclphysb.2013.05.017}
	[\eprint{1303.4475}].

\bibitem{WB2} 
	B.~Wu and L.~Zhao, 
	``Holographic fluid from nonminimally coupled
	scalar-tensor theory of gravity'', 
	\doi{Class.\ Quant.\ Grav.\  {\bf 31}, 105018 (2014)}
  {doi:10.1088/0264-9381/31/10/105018}
	[\eprint{1401.6487}].

\bibitem{Cai:2014ywa}
  R.~G.~Cai, Q.~Yang and Y.~L.~Zhang,
  ``Petrov type I Spacetime and Dual Relativistic Fluids'',
 	\doi{Phys.\ Rev.\ D {\bf 90}, no. 4, 041901 (2014)}
  {doi:10.1103/PhysRevD.90.041901}
  [\eprint{1401.7792}].

\bibitem{HWZ}
  X.~Hao, B.~Wu and L.~Zhao,
  ``Flat space compressible fluid as holographic dual of black hole with
  curved horizon,''
  \doi{JHEP {\bf 1502}, 030 (2015)}
  {doi:10.1007/JHEP02(2015)030}
  [\eprint{1412.8144}].

\bibitem{wxn} 
  W.~J.~Pan, Y.~Tian and X.~N.~Wu,
  ``From Petrov–Einstein–Dilaton–Axion to Navier–Stokes equation in anisotropic model,''
  \doi{Phys.\ Lett.\ B {\bf 752}, 1 (2016)}
  {doi:10.1016/j.physletb.2015.11.008}
  [\eprint{1508.04972}].

\bibitem{Pan:2016ztm} 
  W.~J.~Pan and Y.~C.~Huang,
  ``Fluid/gravity correspondence for massive gravity,''
  \doi{Phys.\ Rev.\ D {\bf 94}, no. 10, 104029 (2016)}
  {doi:10.1103/PhysRevD.94.104029}
  [\eprint{1605.02481}].

\bibitem{Ling:2013kua} 
  Y.~Ling, C.~Niu, Y.~Tian, X.~N.~Wu and W.~Zhang,
  ``Note on the Petrov-like boundary condition at finite cutoff surface in gravity/fluid duality,''
  \doi{Phys.\ Rev.\ D {\bf 90}, no. 4, 043525 (2014)}
  {doi:10.1103/PhysRevD.90.043525}
  [\eprint{1306.5633}].

\bibitem{Bjorken:1982qr} 
  J.~D.~Bjorken,
  ``Highly Relativistic Nucleus-Nucleus Collisions: The Central Rapidity Region,''
  \doi{Phys.\ Rev.\ D {\bf 27}, 140 (1983).}
  {doi:10.1103/PhysRevD.27.140}

\bibitem{Janik:2006ft} 
  R.~A.~Janik,
  ``Viscous plasma evolution from gravity using AdS/CFT,''
  \doi{Phys.\ Rev.\ Lett.\  {\bf 98}, 022302 (2007)}
  {doi:10.1103/PhysRevLett.98.022302}
  [\eprint{hep-th/0610144}].

\bibitem{Chesler:2008hg} 
  P.~M.~Chesler and L.~G.~Yaffe,
  ``Horizon formation and far-from-equilibrium isotropization in supersymmetric Yang-Mills plasma,''
  \doi{Phys.\ Rev.\ Lett.\  {\bf 102}, 211601 (2009)}
  {doi:10.1103/PhysRevLett.102.211601}
  [\eprint{0812.2053}].

\bibitem{Chesler:2010bi} 
  P.~M.~Chesler and L.~G.~Yaffe,
  ``Holography and colliding gravitational shock waves in asymptotically $AdS_5$ spacetime,''
  \doi{Phys.\ Rev.\ Lett.\  {\bf 106}, 021601 (2011)}
  {doi:10.1103/PhysRevLett.106.021601}
  [\eprint{1011.3562}].


\bibitem{Balasubramanian:2011ur} 
  V.~Balasubramanian, A.~Bernamonti, J.~de Boer,
  N.~Copland, B.~Craps, E.~Keski-Vakkuri, B.~Müller,
  A.~Schäfer, M.~Shigemori, W.~Staessens,
  ``Holographic Thermalization,''
  \doi{Phys.\ Rev.\ D {\bf 84}, 026010 (2011)}
  {doi:10.1103/PhysRevD.84.026010}
  [\eprint{1103.2683}].

\bibitem{AbajoArrastia:2010yt} 
  J.~Abajo-Arrastia, J.~Aparicio and E.~Lopez,
  ``Holographic Evolution of Entanglement Entropy,''
  \doi{JHEP {\bf 1011}, 149 (2010)}
  {doi:10.1007/JHEP11(2010)149}
  [\eprint{1006.4090}].


\bibitem{Bhattacharyya:2008ji} 
  S.~Bhattacharyya, R.~Loganayagam, S.~Minwalla, 
  S.~Nampuri, S.~P.~Trivedi and S.~R.~Wadia,
  ``Forced Fluid Dynamics from Gravity,''
  \doi{JHEP {\bf 0902}, 018 (2009)}
  {doi:10.1088/1126-6708/2009/02/018}
  [\eprint{0806.0006}].

\bibitem{Cai:2012vr} 
  R.~G.~Cai, L.~Li, Z.~Y.~Nie and Y.~L.~Zhang,
  ``Holographic Forced Fluid Dynamics in Non-relativistic Limit,''
  \doi{Nucl.\ Phys.\ B {\bf 864}, 260 (2012)}
  {doi:10.1016/j.nuclphysb.2012.06.014}
  [\eprint{1202.4091}].

\bibitem{Hayward:1993ph} 
  S.~A.~Hayward,
  ``Quasilocal gravitational energy,''
  \doi{Phys.\ Rev.\ D {\bf 49}, 831 (1994)}
  {doi:10.1103/PhysRevD.49.831}
  [\eprint{gr-qc/9303030}].

\bibitem{Hu:2015xva} 
  Y.~P.~Hu and H.~Zhang,
  ``Misner-Sharp Mass and the Unified First Law in Massive Gravity,''
  \doi{Phys.\ Rev.\ D {\bf 92}, no. 2, 024006 (2015)}
  {doi:10.1103/PhysRevD.92.024006}
  [\eprint{1502.00069}].

\bibitem{Moskalets:2014hoa} 
  T.~Moskalets and A.~Nurmagambetov,
  ``Liouville mode in Gauge/Gravity Duality,''
  \doi{Eur.\ Phys.\ J.\ C {\bf 75}, no. 11, 551 (2015)}
  {doi:10.1140/epjc/s10052-015-3772-3}
  [\eprint{1409.4186}].

\bibitem{Hao:2015pal} 
  X.~Hao, B.~Wu and L.~Zhao,
  ``Fluids and vortex from constrained fluctuations around C-metric black hole,''
  [\eprint{1511.08281}].
\end{thebibliography}
\end{document}